\documentclass[prb,amsmath,superscriptaddress,citeautoscript,twocolumn,showpacs,floatfix]{revtex4}
\usepackage{graphicx}

\usepackage{amssymb}
\usepackage{color}
\usepackage{epsfig}

\def\sig{{\mbox{\boldmath{$\sigma$}}}}

\def\Omeg{{\mbox{\boldmath{$\Omega$}}}}

\def\rh{{\mbox{\boldmath{$\rho$}}}}
\def\b0{{\bf{0}}}
\def\Del{{\mbox{\boldmath{$\Delta$}}}}

\begin{document}

\title{Retrieving qubit information despite decoherence}
\author{Amnon Aharony}
\email{aaharony@bgu.ac.il} \altaffiliation{Also at Tel Aviv
University, Tel Aviv 69978, Israel.}

\affiliation{Department of Physics and the Ilse Katz Center for
Meso- and Nano-Scale Science and Technology, Ben-Gurion
University, Beer Sheva 84105, Israel}

\affiliation{Albert Einstein Minerva Center for Theoretical
Physics, Weizmann Institute of Science, Rehovot 76100, Israel}

\author{Shmuel Gurvitz}
\affiliation{Department of Particle Physics and Astrophysics,
Weizmann Institute of Science, Rehovot 76100, Israel}

\author{Ora Entin-Wohlman}
\altaffiliation{Also at Tel Aviv University, Tel Aviv 69978,
Israel.}

\affiliation{Department of Physics and the Ilse Katz Center for
Meso- and Nano-Scale Science and Technology, Ben-Gurion
University, Beer Sheva 84105, Israel}

\affiliation{Albert Einstein Minerva Center for Theoretical
Physics, Weizmann Institute of Science, Rehovot 76100, Israel}

\author{Sushanta Dattagupta}
\affiliation{ Indian Institute of Science Education and
Research-Kolkata, Mohanpur 741252, India}

\date{\today}
\begin{abstract}

The time evolution of a qubit, consisting of two single-level
quantum dots, is studied in the presence of telegraph noise. The
dots are connected by two tunneling paths, with an Aharonov-Bohm
flux enclosed between them. Under special symmetry conditions,
which can be achieved by tuning gate voltages, there develops
partial decoherence: at long times, the off-diagonal element of
the reduced density matrix (in the basis of the two dot states)
approaches a non-zero value, generating a circulating current
around the loop. The flux dependence of this current contains full
information on the initial quantum state of the qubit, even at
infinite time. Small deviations from this symmetry yield a very
slow exponential decay towards the fully-decoherent limit.
However,  the amplitudes of this decay also contain the full
information on the initial qubit state, measurable  either via the
current or via the occupations of the qubit dots.

\end{abstract}

\maketitle

\section{Introduction and summary}

Quantum computation operates on information  stored in ``qubits",
which are  superpositions of two basic quantum states 
\cite{bennett},
\begin{align}\label{psi}
|\psi^{}_0\rangle=\cos\alpha|1\rangle+e^{i\gamma}\sin\alpha
|2\rangle\ ,
\end{align}
with two real parameters $\alpha$ and $\gamma$. In one realization
of a solid-state qubit, the two basic states represent
single-level quantum dots \cite{divin}, where the superposition
state (\ref{psi}), representing a single electron on the two dots,
may be given as input, or modified by tuning the dot energies
$\epsilon^{}_{1,2}$ and the inter-dot tunneling $J^{}_{12}$. In a
tight-binding language, the Hamiltonian of the qubit is given by
\begin{align}\label{Hs0}
 \mathcal{H}^{}_{\rm q}&=\epsilon^{}_1a^\dagger_1 a^{}_1+\epsilon^{}_2a^\dagger_2 a^{}_2
 -(J^{}_{12}a^\dagger_1 a^{}_2+{\rm h.c.})
 ~,
\end{align}
where $a^\dagger_n$ creates an electron on dot $n$,
$a^\dagger_n|0\rangle\equiv |n\rangle$. The average energy
$\epsilon\equiv(\epsilon^{}_1+\epsilon^{}_2)/2$ does not affect
the dynamics of the qubit,\cite{you} and therefore we set it equal
to zero. The dynamics is then determined by the energy gap
$\Delta=\epsilon^{}_1-\epsilon^{}_2$ and by $J^{}_{12}$,
\begin{align}\label{Hs}
 \mathcal{H}^{}_{\rm q}&=(\Delta/2)(a^\dagger_1 a^{}_1-a^\dagger_2
 a^{}_2)
 -(J^{}_{12}a^\dagger_1 a^{}_2+{\rm h.c.})
 ~.
\end{align}
In the literature on NMR,\cite{slichter,abragam,Allen}
$\mathcal{H}^{}_{\rm q}$ is often written in the equivalent
pseudo-spin form
\begin{align}
\mathcal{H}^{}_{\rm q}={\bf B}\cdot\sig\ ,
\end{align}
where $\sig$ represents the three Pauli matrices and where
$B^{}_z=\Delta/2,~B^{}_x-iB^{}_y=-J^{}_{12}$. Another physical
realization of a qubit involves a superconducting Josephson
junction. \cite{nakamura}

   Clearly, quantum computation requires the
stability of the quantum state stored on each qubit, and therefore
it can be used only while this state remains coherent
\cite{zurek}. Interactions between  qubits and their environment,
including input-output measurement devices, can cause decoherence
which destroys the information stored in the qubits. Therefore, it
is important to study the time evolution of the qubit state in the
presence of the environment. To concentrate on the state of the
qubit, one traces over the states of the environment, ending up
with the $2\times 2$ reduced density matrix of the qubit itself,
$\rho(t)\equiv{\rm Tr}^{}_{\rm
env}[|\Psi(t)\rangle\langle\Psi(t)|]$, where $|\Psi\rangle$ is the
combined state of the qubit and the environment. In many cases,
the coupling to the environment yields {\it full asymptotic
decoherence}, for which the elements of reduced density matrix
approach the {\it fully-mixed state},
\begin{align}\label{mixed}
\rho^{}_{nm}(t\rightarrow\infty)=\delta^{}_{nm}/2\ , \end{align}
independent of which basis is used for the Hilbert space. In these
cases, the information on the initial qubit quantum state is
totally lost. However, in some symmetric cases there exist
decoherence-free subspaces, which decouple from the environment,
so that at least some of the information on the initial quantum
state remains protected.\cite{Lidar,Zanardi}

Here we discuss special cases in which the full information on the
initial qubit state can be retrieved, even after a long time,
despite decoherence.  Since our results rely mainly on symmetry,
we expect them to hold whenever the required symmetry holds,
irrespective of the specific nature of the environment and its
coupling to the qubit. To demonstrate our point we consider the
simplest model for decoherence, where the environment generates a
single parameter which fluctuates randomly between two values, as
in telegraph noise
 \cite{datta}. We thus replace the Hamiltonian of the qubit by
 $\mathcal{H}^{}_{\rm q}\rightarrow
\mathcal{H}^{}_{\rm q}+f(t)\mathcal{V}$, where $f(t)$ jumps
stochastically between $+1$ and $-1$. Indeed, such jumps in $f(t)$
may arise e.g. due to equilibrium or non-equilibrium sources of
noise, e.g. from background (natural) charge
fluctuations\cite{bergli,IT} or to a capacitive coupling to a
current which flows through a (tunable) neighboring
single-electron transistor.\cite{gur2,gur3}

In principle, the Hamiltonian of the coupling between the qubit
and the noise source, $\mathcal{V}$, may involve the same
operators which appear in Eq. (\ref{Hs}):
\begin{align}\label{VV}
 \mathcal{V}=(\zeta^{}_\Delta/2)(a^\dagger_1 a^{}_1-a^\dagger_2
 a^{}_2)
 -(\zeta^{}_{J}a^\dagger_1 a^{}_2+{\rm h.c.})
 ~,
\end{align}
where the $\zeta$'s are fixed coefficients, whose size measures
the coupling between the qubit and the environment. In the special
cases which we discuss, $\mathcal{V}$ commutes with
$\mathcal{H}^{}_{\rm q}$. This requires  specific ratios between
the coefficients in Eq. (\ref{VV}) and those in Eq. (\ref{Hs}).
Such ratios can be achieved experimentally by tuning gate voltages
which control the coefficients in Eq. (\ref{Hs}), or by a careful
placing of the source of the noise relative to the qubit. When
these conditions are obeyed, one can switch to a basis which
diagonalizes $\mathcal{H}^{}_{\rm q}$. In this basis, the diagonal
elements of the reduced density matrix are independent of time,
while the off-diagonal elements decay to zero, reflecting pure
dephasing. Translated
 to the dot basis of the Hilbert space, this implies that
\begin{align}\label{Vav}
&{\rm Tr}^{}_{\rm env}[\langle\Psi(t)|\mathcal{H}^{}_{\rm
q}|\Psi(t)\rangle]\nonumber\\
&\equiv\frac{\Delta}{2} [\rho^{}_{11}(t)-\rho^{}_{22}(t)]-2{\rm
Re}[J^{}_{12}\rho^{}_{21}(t)]={\rm const.}\ ,
\end{align}
independent of time.
 Thus, the density matrix never reaches the fully-mixed state (\ref{mixed}). We
refer to this situation as {\it partial decoherence}. 

The decoherence of qubits due to  telegraph noise has been treated
in several earlier papers \cite{IT,bergli,gur2,gur3,also}. In a
situation where both $\Delta$ and $J^{}_{12}$ are present, and
both are noisy, the qubit's reduced density matrix usually decays
exponentially towards the fully-mixed state.\cite{stamp}
Alternatively, Itakura and Tokura\cite{IT} considered  the special
case without a gap between the dot energies,
 $\Delta=\zeta^{}_\Delta=0$ [cf. Eqs. (\ref{Hs}) and (\ref{VV})],
 and found that when $J^{}_{12}$ and $\zeta^{}_J$ are real,
 then both $\mathcal{H}^{}_{\rm q}$ and $\mathcal{V}$ are symmetric
 under the interchange $1 \leftrightarrow 2$,
 and therefore the `bonding'
 and `anti-bonding' symmetric and antisymmetric states
 $|\pm\rangle=(|1\rangle\pm |2\rangle)/\sqrt{2}$ are eigenstates of
 both. In that case, the
 off-diagonal element $\rho^{}_{+-}$ oscillates and decays to zero
 with the `dephasing time' $T^{}_2$. Below we show that this
 example is a special case of a broad family of systems, all of
 which exhibit partial decoherence [cf. Eq. (\ref{Vav})].

\begin{figure}[h]
\includegraphics[width=3.5cm]{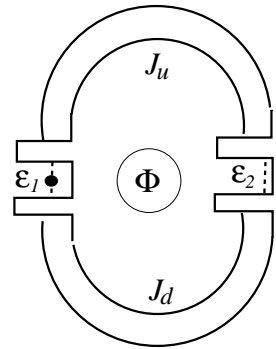}
\caption{The two-dot qubit with tunneling channels $u$ and $d$. }
\label{fig0}
\end{figure}

The procedures proposed below require that $J^{}_{12}$ should be
complex and tunable experimentally.  To achieve this, we connect
the two qubit quantum dots
 via two separate tunneling
channels, with energies $J^{}_u$ and $J^{}_d$ (Fig. \ref{fig0}),
while a magnetic flux $\Phi$ is enclosed between them. Utilizing
gauge invariance, the combined tunneling coupling $J^{}_{12}$
becomes
\begin{align}\label{J12}
J^{}_{12}=J^{}_u+J^{}_d e^{i\phi}\equiv |J^{}_{12}|e^{i\theta}\ ,
\end{align}
with the Aharonov-Bohm phase $\phi=2\pi\Phi/\Phi^{}_0$, where
$\Phi^{}_0=h c/e$ is the flux unit. Both $J^{}_u$ and $J^{}_d$
(which are chosen real) can be tuned via gate voltages, and the
phase $\theta$ can be tuned via the magnetic flux.

To demonstrate the result (\ref{Vav}) we present what we term `the
symmetric case', in which there is no energy gap,
$\Delta=\zeta^{}_\Delta=0$, but $J^{}_{12}$ can still be complex.
The vanishing of $\zeta^{}_\Delta$ is achieved when the source of
noise is located symmetrically relative to the two dots, or when
the correlation length of the noise is larger than the qubit's
size.
 For an arbitrary flux $\Phi$, and for given values of the noise
 coefficients $\zeta^{}_u$ and $\zeta^{}_d$ [defined
 via $J^{}_{u,d}\rightarrow J^{}_{u,d}+f(t)\zeta^{}_{u,d}$],
 we show that an appropriate tuning of either $J^{}_u$ or $J^{}_d$
 (via corresponding gate voltages) suffices to
 bring the system into a special symmetric case, in which its
 reduced density matrix {\it never} reaches the fully-mixed state
 (\ref{mixed}). Instead, at $t\rightarrow\infty$ it approaches the asymptotic limit
\begin{align}
\rho\rightarrow\left(\begin{array}{cc}
 1/2 & e^{i\theta}{\rm Re}[e^{-i\theta}\rho^{}_{12}(0)] \\
 e^{-i\theta}{\rm Re}[e^{-i\theta}\rho^{}_{12}(0)] & 1/2
\end{array}\right)\ ,\label{r11}
\end{align}
where  $\rho^{}_{12}(0)=\langle 1|\psi^{}_0\rangle\langle\psi^{}_0
|2\rangle=e^{-i\gamma}\sin(2\alpha)/2$ [see Eq. (\ref{psi})] and
where  $\theta$ is the phase of the complex $J^{}_{12}$, see Eq.
(\ref{J12}). Equation (\ref{r11}) is a special case of Eq.
(\ref{Vav}), for $\Delta=0$.

The non-zero complex off-diagonal element $\rho^{}_{12}$ generates
a circulating current around the loop, which can be used to
retrieve the qubit's information. The equation of motion for the
number operator $a^\dagger_1 a^{}_1$ is
\begin{align}
\partial^{}_t a^\dagger_1
a^{}_1=\hat{\cal I}^{}_u+\hat{\cal I}^{}_d\ ,
\end{align}
where the operator $\hat{\cal I}^{}_{u}$ and $\hat{\cal I}^{}_{d}$
represent the currents into dot 1 via the channels $u$ and $d$,
respectively:
\begin{align}
\hat{\cal I}^{}_{u}=i[J^{}_{u}a^\dagger_1 a^{}_2-{\rm h.c.}]\ ,\ \
\ \hat{\cal I}^{}_{d}=i[J^{}_{d}e^{i\phi}a^\dagger_1 a^{}_2-{\rm
h.c.}]\ .
\end{align}
Therefore, the net current into site 1 is
\begin{align}\label{netcur}
{\cal I}=\langle\Psi|\hat{\cal I}^{}_u-\hat{\cal
I}^{}_d|\Psi\rangle=\langle\Psi|2\hat{\cal
I}^{}_u-\partial^{}_ta^\dagger_1a^{}_1|\Psi\rangle\ .
\end{align}
In the asymptotic stationary limit we can drop the last term.
Tracing over the environment then yields the conditional average
$\langle\Psi(t)|J^{}_ua^\dagger_1a^{}_2|\Psi(t)\rangle\rightarrow
J^{}_u\rho^{}_{21}$. For the symmetric case, we find below that
averaging over the noise yields
\begin{align}\label{curcur}
{\cal I}&\rightarrow 2J^{}_0{\rm
Im}[\rho^{}_{12}(\infty)]
=J^{}_0\sin(2\alpha)\sin\theta\cos(\theta+\gamma)\
,
\end{align}
where $J^{}_0$ is the average of $J^{}_u+\zeta^{}_uf(t)$ over the
noise. As expected, this current vanishes when $\phi$ is an
integer multiple of $\pi$ [when also $\theta=0$, see Eq.
(\ref{J12})]. However, at non-trivial fluxes ${\cal I}$ is
non-zero, despite decoherence. This current generates an orbital
magnetic moment of the electron circulating the loop. Measuring
its $\phi-$dependence can yield both $\alpha$ and $\gamma$ [see
Eq. (\ref{psi})], namely the full information stored initially!
Unlike the usual equilibrium persistent current, which is an odd
function of the flux (as required by time-reversal
symmetry),\cite{Imry} the current here is neither odd nor even in
the flux. This peculiar flux dependence apparently results from
the averaging over the noise, which breaks time-reversal symmetry.

Following our analysis of the symmetric case, we also consider
small deviations from this symmetry, and find that
 such deviations lead asymptotically to the {\it fully-mixed
state}, as also found e. g. in Refs.$\
$\onlinecite{IT}--\onlinecite{gur2} and in references given there.
However, for small deviations from symmetry there is a distinct
separation of time scales.  After a transient oscillatory stage,
the elements of the density matrix develop a very slow simple
exponential decay towards the asymptotic fully-mixed state. These
slowly decaying terms also include a non-zero difference in the
dot occupations, $z=\rho^{}_{11}-\rho^{}_{22}$ (which approached
zero relatively quickly in the symmetric case). This difference
can also be measured experimentally. We show that the amplitudes
of these exponential terms also contain the full information on
the initial state of the qubit. This information can therefore be
extracted even after the transient stage.

The plan of the paper is as follows.  The formalism for the
telegraph noise is reviewed in Sec. II.
 Section III then presents several physical environments which
 can generate telegraph noise of the kind discussed here. The
 general conditions for partial decoherence, and the example of
 the symmetric case, are presented in Sec. IV.  In Sec. V we then introduce deviations from
symmetry, and  Sec. VI contains a discussion of our results.

\section{Telegraph noise}

A treatment of the equation of motion with the underlying
stochasticity in $f(t)$ can be found in the literature on the
theory of lineshapes \cite{slichter,10}. Here we follow Blume
\cite{11}, and average the density matrix $\rho(t)$ over the
histories of the stochastic noise, under the condition that at
time $t$ the random function $f(t)$ has the values $b=1$ or $-1$.
We then define a $2-$component vector (denoted by bold letters)
$\rh(t)$, such that its $b-$th component represents this
conditional average $\rho(t,b)$.
At the end
one may average over the stochastic process,
\begin{align}\label{av}
 \rho(t)=\sum_{b=\pm 1} {\rho}^{}(t,b) ~.
\end{align}

The function $f(t)$ follows a Markov process:\cite{12} it jumps
randomly from $1$
 to $-1$ (or from $-1$ to $1$) with the rate
$w^{}_{-+}$ (or $w^{}_{+-}$). These jumps in $f(t)$ result from a
contact with some noise source. The noise distribution is
characterized by the probabilities $p^{}_{\pm}$ to find $f(t)$ at
the values $\pm 1$. Detailed balance then implies the relation
$p^{}_{-}w^{}_{+-}=p^{}_{+}w^{}_{-+}$, and therefore the jump
rates can be written as \begin{align} w^{}_{\pm \mp}=\lambda
p^{}_{\pm}\ , \label{wp} \end{align} where
 $ \lambda =w^{}_{+-} + w^{}_{-+}$ represents the inverse time associated with the noise.

Our main purpose here is to calculate the time evolution of $\rh$.
The equations of motion for the conditional averages $\rho(t,b)$
are
\begin{align}
\partial_t\rho(t,b)&=-i\bigl [\mathcal{H}{}_{\rm q}+ b \mathcal{V},\rho(t,b)\bigr ]\nonumber\\
&+w^{}_{b,-b}\rho(t,-b)-w^{}_{-b,b}\rho(t,b)~,\label{dtrho}
\end{align}
where we use $\hbar=1$ throughout. The first term on the
right-hand side applies if $f(t)$ remains unchanged at time $t$
(i.e. stays equal to $b$). In this case, the time evolution of the
density matrix proceeds with the Liouville operator which
corresponds to the original Hamiltonian, with $f(t)=b$. The last
two terms arise if $f(t)$ flips exactly at time $t$, either from
$-b$ to $b$ (second term) or from $b$ to $-b$ (last term).

Each element of the $2\times 2$ reduced density matrix now becomes
a $2-$component vector, $\rh^{}_{nm}$, and  Eq. (\ref{dtrho}) can
be written in matrix form,
\begin{align}\label{eom}
i({\bf I}\partial^{}_t-{\bf W})\rh^{}_{nm}=\Del^{}_{nm}\rh^{}_{nm}
-\sum_{\ell}\bigl ({\bf J}^{}_{n\ell}\rh^{}_{\ell
m}-\rh^{}_{n\ell}{\bf J}^{}_{\ell m}\bigr )~,
\end{align}
with $n,m=1,2$. Here, each parameter in the Hamiltonian
$\mathcal{H}{}_{\rm q}+ b \mathcal{V}$ is replaced by a diagonal
$2\times 2$ matrix. For our specific two-dot system,
$\Del^{}_{nm}={\bf J}^{}_{nm}=0$ for $n=m$ while
$\Del^{}_{12}\equiv\Del\equiv \Delta{\bf
I}+\zeta^{}_\Delta\sig^{}_z$ represents the energy gap variable
and $J^{}_{12}\rightarrow {\bf J}^{}_{12}\equiv J^{}_{12}{\bf
I}+\zeta^{}_J\sig^{}_z$ represents the hopping matrix element
(${\bf I}$ is the $2\times 2$ unit matrix). The relaxation matrix
${\bf W}$ consists of the stochastic hopping probabilities of the
noise,
\begin{align}\label{WW}
{\bf W}=\left(\begin{array}{cc}
 -w^{}_{-+} & w^{}_{+-} \\
 w^{}_{-+} & -w^{}_{+-}
\end{array}\right)\equiv \lambda({\bf T}-{\bf I})
\end{align}
 [see Eq. (\ref{wp})], where
 \begin{align}\label{TT}
{\bf T}\equiv \left(\begin{array}{cc}
 p^{}_+ & p^{}_+ \\
 p^{}_- & p^{}_-
\end{array}\right)\equiv[{\bf I} + \sig^{}_x + \Delta p(\sig^{}_z + i\sig^{}_y)]/2
\end{align}
and $\Delta p\equiv p^{}_+-p^{}_-$.

The matrix ${\bf W}$ determines the time evolution of the matrix
${\bf P}(t)$, where $(b|{\bf P}(t)|a)$ is the probability for the
stochastic variable to start at $t=0$ with the value $a$ and end
at $t>0$ with the value $b$: ${\bf P}$ obeys the equation
$\partial^{}_t{\bf P}(t)={\bf WP}(t)$, and therefore
\begin{align}\label{Pt}
{\bf P}(t)=e^{{\bf W}t}={\bf T}+\bigl ({\bf I}-{\bf T}\bigr
)e^{-\lambda t}\ ,
\end{align}
where we have used the identity ${\bf T}^2={\bf T}$. At infinite
time, ${\bf P}$ approaches ${\bf T}$, and thus $(b|{\bf
P}(\infty)|a)=p^{}_b$ ($a,b=\pm$), independent of the initial
value $a$. Below we use ${\bf P}(t)$ for some of the solutions for
the density matrix. It is customary to characterize the noise by
its spectral function, defined via
\begin{align}\label{spec}
S(\omega )=2{\rm Re}\int_0^\infty \sum_{b,b'}p^{}_{b'}~(b-\bar
b)(b|{\bf P}(t)|b')(b'-\bar b) e^{i\omega t}dt\ ,
\end{align}
where $\bar b=\sum_b p_b\, b=\Delta p$. Using Eq.~(\ref{Pt}) one
finds
\begin{align}
S(\omega )=\frac{8w_{+-}^{}w_{-+}^{}}{\lambda
(\omega^2+\lambda^2)}\equiv \frac{8\lambda
p^{}_+p^{}_-}{\omega^2+\lambda^2}\ . \label{noisesp}
\end{align}
Below we relate various decay times with special values of
$S(\omega)$.

A convenient way to solve Eqs. (\ref{eom}) is by employing the
Laplace transform,
\begin{align}
\widetilde{\rh}(s)=\int_{0}^{\infty}dt e^{-st}\rh(t)\ .
\end{align}
The equations of motion (\ref{eom}) then become
\begin{align}\label{eomL}
({\bf I}s-{\bf W})\widetilde{\rh}^{}_{nm}=\rh^{}_{nm}(0)-i{\Del}^{}_{nm}\widetilde{\rh}^{}_{nm}\nonumber\\
+i\sum_{\ell}\bigl ({\bf J}^{}_{n\ell}\widetilde{\rh}^{}_{\ell
m}-\widetilde{\rh}^{}_{n\ell}{\bf J}^{}_{\ell m}\bigr )~.
\end{align}
Since the initial values of the density matrix,
$\rho^{}_{nm}(0)\equiv\langle
n|\psi^{}_0\rangle\langle\psi^{}_0|m\rangle$, do not depend on the
stochastic noise, we assume the latter to be in its steady state,
and identify the initial vector $\rh^{}_{nm}(0)$ with
\begin{align}\label{rho0}
\rh^{}_{nm}(0)\equiv \rho^{}_{nm}(0){\bf p}^{}_0,\ \ \ {\bf p}^{}_0\equiv\left(\begin{array}{c}
 p^{}_+ \\
 p^{}_-
\end{array}\right)\ .
\end{align}
Note that ${\bf p}^{}_0$ is an eigenstate of ${\bf T}$, with
eigenvalue $1$, and therefore ${\bf P}(t){\bf p}^{}_0={\bf
p}^{}_0$, so that ${\bf p}^{}_0$ corresponds to the steady state
of the stochastic noise.

One major issue in this paper concerns the asymptotic limit of the
density matrix, at long times. Without noise, the eigenvalues of
$\mathcal{H}_{\rm q}$ are $\pm\Omega/2$, with eigenstates
$|\pm\rangle$. In the basis of these eigenstates,
$\rho^{}_{\pm\pm}$ remain constant in time, while
$\rho^{}_{+-}(t)=e^{-i\Omega t}\rho^{}_{+-}(0)$, with the Rabi
frequency
\begin{align}\label{Omega}
\Omega=\sqrt{\Delta^2+4|J^{}_{12}|^2}\ . \end{align}
 In the
presence of noise, the density matrix often approaches a
stationary state, so that $\partial^{}_t\rh=0$. This is indeed the
case in our analysis. One can then find the stationary state by
solving the homogeneous linear set of equations (\ref{eom}) in
$\rh^{}_{nm}$. Alternatively, one can use Eqs. (\ref{eomL}),
together with the identity
\begin{align}
\lim^{}_{t\rightarrow\infty}\rh(t)=\lim^{}_{s\rightarrow
0}s\widetilde{\rh}(s)\ .\label{lim}\end{align}

\section{Physical models yielding telegraph noise}

The reduced density matrix is obtained by solving  the equations
of motion for the joint density matrix of the qubit and the
environment, and then tracing over the environment degrees of
freedom. This procedure is quite complicated when the
time-dependence of the environment degrees of freedom is
influenced by those of the qubit.  This influence is called `back
action'. Neglecting this back action implies that one can
calculate the time dependence of the environment degrees of
freedom separately, independent of the qubit states. In the
simplest model discussed here, the qubit couples to the
environment only via one degree of freedom, which is represented
by its time dependent value $f(t)$.  Here we review three examples
of models which have been treated in the literature.

Itakura and Tokura \cite{IT} already reviewed the literature on
background charge fluctuations (see also Refs.$\ $
\onlinecite{bergli} and$\ $\onlinecite{also}). In that case, a
single impurity near the qubit is either occupied by an electron
or empty, with probabilities $p^{}_+$ and $p^{}_-$ and with
hopping rates given by Eq. (\ref{wp}), in which $\lambda$ is
proportional to $e^{-\Delta E/k^{}_B T}$ and $\Delta E$ is the
activation energy of the impurity (we asume a large Coulomb
blockade, preventing double occupancy). Neglecting the back action
of the qubit onto the impurity, this model reduces to the
classical telegraph noise one, in which the impurity-qubit
coupling generates different coefficients in the intra-qubit
Hamiltonian for each state of the impurity.

Another possible model concerns a two-level system with an energy
gap $\Delta E$ (e.g. a double potential well created by two
neighboring impurities\cite{bergli}), which can be represented by
a pseudo-spin $1/2$. At equilibrium  with a heat bath at
temperature $T$, the occupation probabilities obey a Boltzmann
distribution
 $p^{}_+=1-p^{}_-=[1+e^{-\Delta E/k^{}_BT}]^{-1}$. Each
 state of this pseudo-spin generates different values for the
 coupling parameters within the qubit Hamiltonian, again yielding the
 telegraph noise picture.

It is usually not easy to justify the neglect of the back actions
in the above two examples. However, as argued by Galperin {\it et
al.},\cite{galp} back action may be ignored when the dynamics of
the fluctuating background charge or the two-level system is
governed by its coupling to a thermalizing heat bath, which is
much stronger than its coupling to the qubit. The telegraph noise
model is also justified in the limit of a very high temperature of
this heat bath.\cite{also}

Here we concentrate on yet another example, in which a current
between a left and a right reservoirs $L$ and $R$ respectively
(held at chemical potentials $\mu^{}_L>\mu^{}_R$) flows through  a
single electron transistor (SET), located near the qubit. Unlike
the above two examples, here the fluctuator is not at equilibrium.
The states of the environment (SET plus reservoirs) include states
in which an arbitrary number of electrons have moved between the
two reservoirs, while the SET can be (singly) occupied or empty.
The quantum equations of motion for the density matrix of the
combined qubit-SET system were analyzed in Refs.~
\onlinecite{gur2} and$\ $\onlinecite{gur3}. When the bias voltage
$\mu^{}_L-\mu^{}_R$ is much larger than any other energy in the
problem, these authors traced over  the environment states, and
obtained equations of motion for the reduced qubit density matrix,
which are equivalent to our Eqs. (\ref{dtrho}). In these
equations, the rate of an electron entering the SET from the left
reservoir, $\Gamma^{}_L$, was identified with $w_{+-}^{}$, and the
rate of an electron leaving the SET to the right, $\Gamma^{}_R$,
was identified with $w_{-+}^{}$. Here, $\Gamma^{}_{L}$
($\Gamma^{}_R$ is the partial width of the SET level, caused by
its coupling to the $L$ ($R$) reservoir.

When the SET is placed near the qubit, the electron on the latter
feels an additional Coulomb potential generated whenever the SET
dot is occupied. Denoting the creation operator of an electron on
the SET by $c^\dagger_0$,  the coupling between the SET and the
qubit is given by
\begin{align}\label{Hint}
\mathcal{H}^{}_{\rm int}&=c_0^\dagger c^{}_0\bigl
(U^{}_1a_1^\dagger a^{}_1+U^{}_2a_2^\dagger
a^{}_2-[U^{}_Ja_1^\dagger a^{}_2+{\rm h.c.}]\bigr )\ .
\end{align}
The energy $U^{}_J$ represents a sum of two matrix elements,
associated with the effect of the SET on the hopping between the
qubit dots. Assuming the geometry of Fig. 1, and using the same
gauge choice as in Eq. (\ref{J12}), these matrix elements can be
written as
\begin{align}\label{UJ}
U^{}_J=U^{}_u +U^{}_d e^{i\phi}\ ,
\end{align}
with real $U^{}_u$ and $U^{}_d$. Using the conclusions of Refs.~
\onlinecite{gur2} and$\ $\onlinecite{gur3}, one may replace
$c^\dagger_0c^{}_0$ in Eq. (\ref{Hint}) by a c-number,
 $[1+f(t)]/2$. Absorbing the
time-independent part in $\mathcal{H}^{}_{\rm q}$ then yields
\begin{align}\label{zeta}
\zeta^{}_\Delta=(U^{}_1-U^{}_2)/2\ ,\ \ \ \
\zeta^{}_{u,d}=U^{}_{u,d}/2\ .
\end{align}
 These parameters clearly depend on the relative location of the
 SET with respect the two dots and the two tunneling paths.


\section{ partial decoherence}

\subsection{General conditions for partial decoherence}

Here we  show that a system develops partial decoherence, i.e.
does not approach the fully-mixed state (\ref{mixed}), whenever
$\mathcal{H}^{}_{\rm q}$ and $\mathcal{V}$ [Eqs. (\ref{Hs}) and
(\ref{VV})] commute with each other, and discuss the conditions
for this to happen. The commutator of these operators is given by
\begin{align}
[\mathcal{H}^{}_{\rm
q},\mathcal{V}]&=[(\Delta\zeta^{*}_J-J^{*}_{12}\zeta^{}_\Delta)a^\dagger_1a^{}_2-{\rm
h.c.}]\nonumber\\
&+(J^{}_{12}\zeta^*_J-J^{*}_{12}\zeta^{}_J)(a^\dagger_1a^{}_1-a^\dagger_2a^{}_2)\
.
\end{align}
This commutator vanishes whenever
\begin{align}\label{cond1}
\Delta\zeta^{}_J-J^{}_{12}\zeta^{}_\Delta=
J^{}_{12}\zeta^{*}_J-J^*_{12}\zeta^{}_J=0 \ ,
\end{align}
namely
\begin{align}\label{cond}
&\frac{\zeta^{}_\Delta}{\Delta}=\frac{\zeta^{}_J}{J^{}_{12}}=\frac{\zeta^{*}_J}{J^{*}_{12}}\equiv
K\ \ \ {\rm when}\  \Delta\ne 0\ ,\nonumber\\
&\frac{\zeta^{}_J}{J^{}_{12}}=\frac{\zeta^{*}_J}{J^{*}_{12}}\equiv
K\ \ {\rm and}\  \zeta^{}_\Delta= 0\ \ \ {\rm when}\  \Delta= 0\ ,
\end{align}
where $K$ is a fixed real number. It should be emphasized that the
conditions (\ref{cond}) apply for {\it any environment}, when f(t)
is replaced by an operator acting on the environment (e.g.
$c^\dagger_0c^{}_0$ in the previous section), and are not
restricted to the telegraph noise example discussed below.

Equation (\ref{cond}) requires that $\zeta^{}_J/J^{}_{12}$ should
be a real number. Using the gauge choice (\ref{J12}) also for the
$\zeta$'s, this would require
$\zeta^{}_u/J^{}_u=\zeta^{}_d/J^{}_d$, or $q^{}_\zeta=q^{}_J$,
where \begin{align}\label{qq}
 q^{}_J\equiv J^{}_u/J^{}_d\ ,\ \ \
q^{}_\zeta\equiv \zeta^{}_u/\zeta^{}_d\ . \end{align}
 Since both
$J^{}_u$ and $J^{}_d$ can be tuned by gate voltages, this
condition can be achieved experimentally.  Alternatively, as
mentioned in the Introduction, one can imagine conditions under
which $\zeta^{}_\Delta=0$, e.g. when the noise source is placed
symmetrically with respect to the two dots. In that case  $\Delta$
can be tuned to zero, and  Eq. (\ref{cond1}) may still hold.

 In both of these
cases, $\mathcal{H}^{}_{\rm q}$ and $\mathcal{V}$ can be
diagonalized simultaneously. The corresponding eigenvalues are
then
\begin{align}\label{eigen}
 \epsilon^{}_\pm=\pm\Omega/2\ ,\ \
\mathcal{V}^{}_{\pm\pm}=\pm \zeta^{}_\Omega/2=\pm
\sqrt{\zeta_\Delta^2/4+|\zeta^{}_J|^2}\ ,
\end{align}
see Eq. (\ref{Omega}). Denoting the corresponding common
eigenstates by $|+\rangle$ and $|-\rangle$ (do not confuse with
the $\pm$ states of the fluctuator), the equations of motion
(\ref{eomL}) in this new basis become
\begin{align}\label{eom2}
&({\bf I}s-{\bf
W})\widetilde{\rh}^{}_{\pm\pm}=\rh^{}_{\pm\pm}(0)\ ,\nonumber\\
&({\bf I}s-{\bf W})\widetilde{\rh}^{}_{+-}=\rh^{}_{+-}(0)-i\Omeg
\widetilde{\rh}^{}_{+-}\ ,
\end{align}
with $\Omeg=\Omega{\bf I}+\zeta^{}_\Omega\sig^{}_z$.

 The solution
for the diagonal matrix elements is
\begin{align}
\widetilde{\rh}^{}_{\pm\pm}=[{\bf I}s-{\bf
W}]^{-1}\rh^{}_{\pm\pm}(0)\equiv\widetilde{\bf
P}\rh^{}_{\pm\pm}(0)\ , \end{align} where $\widetilde{\bf P}$ is
the Laplace transform of ${\bf P}(t)$, and therefore
\begin{align}\label{rpp}
\rh^{}_{\pm\pm}(t)={\bf P}(t)\rh^{}_{\pm\pm}(0)\equiv
\rh^{}_{\pm\pm}(0)=\rho^{}_{\pm\pm}(0){\bf p}^{}_0\ ,
\end{align}
where we have used the explicit expressions (\ref{Pt})  and
(\ref{rho0}). The components of these vectors are
$\rho^{}_{\pm\pm}(t,b)=p^{}_b\rho^{}_{\pm\pm}(0)$, and the
 factor $p^{}_b$ represents the probability of
finding the stochastic noise at the state $b$. Averaging over this
noise [Eq. (\ref{av})], we find that
$\rho^{}_{\pm\pm}(t)=\rho^{}_{\pm\pm}(0)$. Thus, the diagonal
elements of the reduced density matrix do {\it not} approach the
fully-mixed limit (\ref{mixed}), implying partial decoherence.
Note that these matrix elements are linear combinations of the
original density matrix elements, with coefficients which involve
the mapping from $|1,2\rangle$ to $|+,-\rangle$. In fact, one can
repeat the above procedure for any operator which commutes with
both $\mathcal{H}^{}_{\rm q}$ and $\mathcal{V}$, see Eq.
(\ref{Vav}).

The second equation (\ref{eom2}) yields
\begin{align}\label{rhpm}
\widetilde{\rh}^{}_{\pm\mp}=\widetilde{\bf
F}^{}_\pm\rh^{}_{\pm\mp}(0)\equiv \widetilde{\bf
F}^{}_\pm\rho^{}_{\pm\mp}(0){\bf p}^{}_0\ ,
\end{align}
where
\begin{align}\label{FF}
&\widetilde{\bf F}^{}_\pm\equiv[s{\bf I}-{\bf W}\pm i\Omeg]^{-1}\nonumber\\
&=\frac{(2s\pm 2i\Omega+\lambda){\bf I}+\lambda[\sig^{}_x+\Delta
p(\sig^{}_z+i\sig^{}_y)]\mp 2i\zeta^{}_\Omega\sig^{}_z}{2[(s\pm
i\Omega)(s \pm i\Omega+\lambda)\pm i\lambda\zeta^{}_\Omega\Delta
p+\zeta_\Omega^2]}\ .
\end{align}
Here we have used Eqs. (\ref{WW}) and (\ref{TT}). Averaging over
the noise [Eq. (\ref{av})], $\rho^{}_{+-}$ is obtained noting that
the average of $\sig^{}_x{\bf p}^{}_0$ is equal to 1, while the
averages of $-i\sig^{}_y{\bf p}^{}_0$ and $\sig^{}_z{\bf p}^{}_0$
are equal to $\Delta p$. The inverse Laplace transform then yields
\begin{align}\label{rpm}
&\rho^{}_{+-}(t)=\bigl
(A^{}_+e^{\alpha^{}_+t}+A^{}_-e^{\alpha^{}_-t}\bigr
)\rho^{}_{+-}(0)\
,\nonumber\\
&A^{}_\pm=\frac{\pm(\lambda-2i\zeta^{}_\Omega\Delta
p)+\sqrt{\lambda^2-4\zeta_\Omega^2-4i\lambda\zeta^{}_\Omega\Delta
p}}{2\sqrt{\lambda^2-4\zeta_\Omega^2-4i\lambda\zeta^{}_\Omega\Delta
p}}\ ,
\end{align}
where
\begin{align}\label{alsym}
2\alpha^{}_{\pm}=-2i\Omega-\lambda\pm\sqrt{\lambda^2-4\zeta_\Omega^2-4i\lambda\zeta^{}_\Omega\Delta
p}\ . \end{align} Therefore, $\rh^{}_{+-}(t)$ oscillates and
decays to zero. This asymptotic limit can also be obtained using
Eq. (\ref{lim}). The real parts of $-\alpha^{}_{\pm}$ represent
two decay rates. For weak coupling between the qubit and the
environment, $|\zeta|\ll \lambda$, the shorter time (associated
with $\alpha^{}_-$) is of order $1/\lambda$, the typical time
between the fluctuator jumps. To leading order in $\zeta/\lambda$,
the longer relaxation time $\tau$ is approximately given by
\begin{align}\label{tau}
\tau^{-1}=-{\rm Re}[\alpha^{}_+]\approx
4\zeta_\Omega^2p^{}_+p^{}_-/\lambda\equiv \zeta_\Omega^2S(0)/2
\end{align} [see Eq. (\ref{noisesp})]. This relation with the
zero-frequency noise spectrum coincides with the well-known
`dephasing' time $T^{}_2\equiv 2\tau$, generated by fluctuations
of the off-diagonal coupling between the two energy states
\cite{gur2,slichter,T2}. Equation (\ref{alsym})  is also the same
as that found (using a different method) by Itakura and
Tokura\cite{IT}, for the special case when
$\Delta=\zeta^{}_\Delta={\rm Im}[J^{}_{12}]={\rm
Im}[\zeta^{}_J]=0$.

In the NMR terminology one distinguishes between purely dephasing
noise, which causes the decay of the off-diagonal element of the
reduced density matrix (in the relevant basis), and purely
relaxaional noise, associated with the decay of the diagonal
matrix elements towards $1/2$. These two types of noise are
associated with the decay times $T^{}_2$ and $T^{}_1$,
respectively.\cite{slichter,abragam}  The model discussed in this
section describes a purely dephasing mechanism, namely
$T^{}_1\rightarrow\infty$. Similar models have been treated in
connection with a coupling to a continuum of phonon
modes.\cite{viola} The result (\ref{tau}) is the same as that
obtained for white noise, when $S(\omega)$ is independent of
$\omega$. This result corresponds to the `motional narrowing
limit' in NMR, when the rate of the noise fluctuation $\lambda$ is
the shortest time in the problem.\cite{abragam}

In the basis of the Hamiltonian eigenstates $|\pm\rangle$, the
off-diagonal element $\rho^{}_{+-}$ decays to zero, and its
magnitude has been used to quantify decoherence.\cite{IT} However,
this decay is basis-dependent:
 in the original basis of the qubit dot states the off-diagonal
matrix elements approach non-zero values. This is true whenever
one encounters partial decoherence. Therefore, using off-diagonal
elements to characterize decoherence may be misleading. It is
better to quantify decoherence  via a basis-independent measure,
e.g. $[1-{\rm Tr}\rho^2]\equiv
2(\rho^{}_{11}\rho^{}_{22}-|\rho^{}_{12}|^2)$. This quantity
approaches $1/2$ for the fully-mixed state, but is larger than
$1/2$ for partial decoherence.

\subsection{Symmetric case}

Equation (\ref{cond}) refers to  two cases: either the Hamiltonian
contains an energy gap $\Delta$, and an associated noise parameter
$\zeta^{}_\Delta$, or both of these variables vanish. Since the
results are qualitatively the same in both cases, we  present
explicit expressions for the simpler case
$\Delta=\zeta^{}_\Delta=0$.  In this case,
$\Omega=2|J^{}_{12}|,~\zeta^{}_\Omega=2|\zeta^{}_J|$, and the two
common eigenstates are easily identified as
\begin{align}\label{bab}
|\pm\rangle=(|1\rangle\mp e^{-i\theta}|2\rangle)/\sqrt{2}\ ,
\end{align}
where $\theta$ is defined in Eq. (\ref{J12}) (and simultaneously
$\zeta^{}_J=|\zeta^{}_J|e^{i\theta}$, since $\zeta^{}_J/J^{}_{12}$
is real).\cite{sym} Substituting $J^{}_{12}$ from Eq. (\ref{J12})
then yields
\begin{align}\label{tantheta}
\tan\theta=\frac{\sin\phi}{q^{}_J+\cos\phi}\  \end{align} with
$q_J=q_\zeta$ given in Eq. (\ref{qq}).

 We can now use the results
from the previous subsection. For that, we need to map the reduced
density matrix from our original basis $\{1,2\}$ to the
`bonding--antibonding' basis (\ref{bab}) and back. Substituting
the initial conditions
\begin{align}\label{map}
&Z(0)\equiv \rho^{}_{++}(0)-\rho^{}_{--}(0)=2{\rm Re}[e^{-i\theta}\rho^{}_{12}(0)]\ ,\nonumber\\
&\rho^{}_{+-}(0)=[\rho^{}_{11}(0)-\rho^{}_{22}(0)]/2+i{\rm
Im}[e^{-i\theta}\rho^{}_{12}(0)]\
\end{align}
into Eqs. (\ref{rpp}) and (\ref{rpm}) yields $\rho^{}_{\pm\pm}(t)$
and $\rho^{}_{+-}(t)$, and the relations
\begin{align}\label{rho12}
&z(t) \equiv \rho^{}_{11}(t)-\rho^{}_{22}(t)=2{\rm Re}[\rho^{}_{+-}(t)]\ ,\nonumber\\
&\rho^{}_{12}(t)=e^{i\theta}\bigl (i{\rm
Im}[\rho^{}_{+-}(t)]-[\rho^{}_{++}(t)-\rho^{}_{--}(t)]/2\bigr )
\end{align}
yield the reduced density matrix in the original basis. It is now
easy to check that at long times the diagonal elements
$\rho^{}_{11}$ and $\rho^{}_{22}$ approach $1/2$, but the
off-diagonal element approaches
$\rho^{}_{12}(t\rightarrow\infty)\rightarrow e^{i\theta}{\rm
Re}[e^{-i\theta}\rho^{}_{12}(0)]$, as in Eq. (\ref{r11}).

\begin{widetext}

\begin{figure}[ht]
\includegraphics[width=5.8cm]{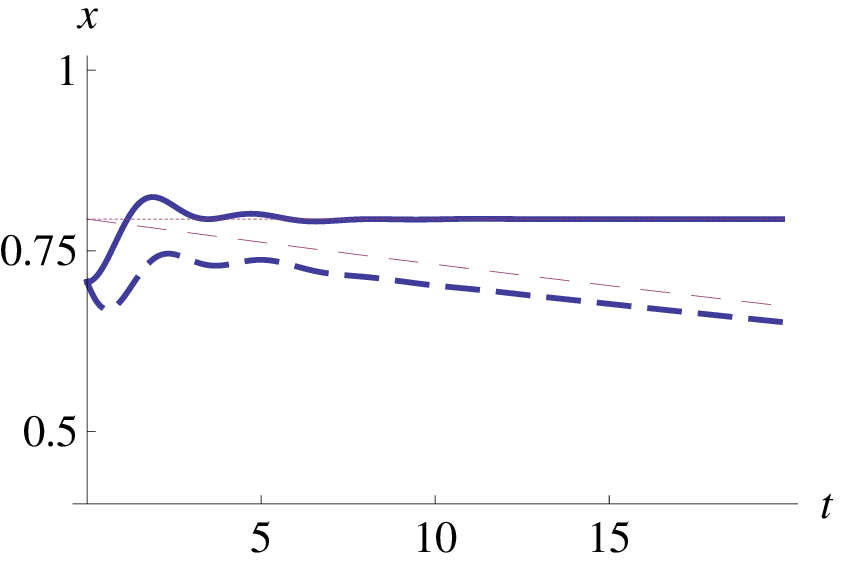}\ \
\includegraphics[width=5.8cm]{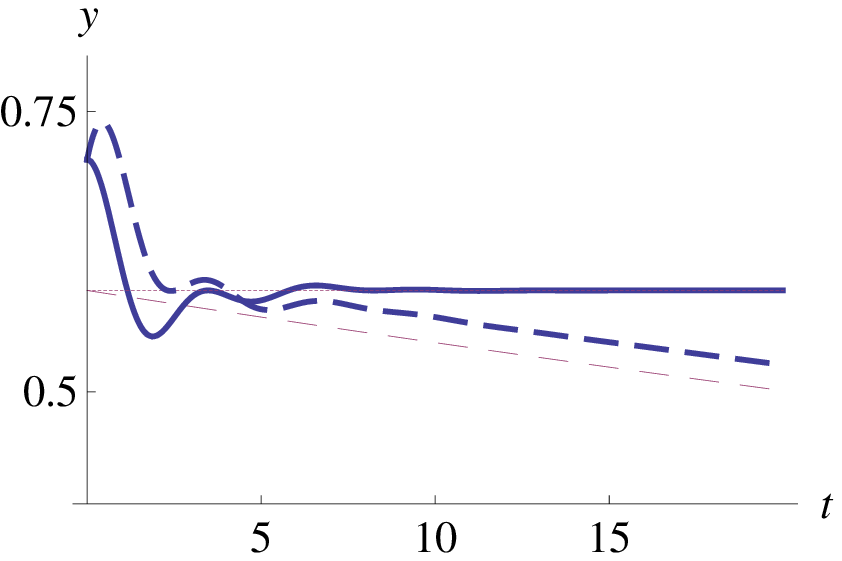}\ \ \
\includegraphics[width=5.8cm]{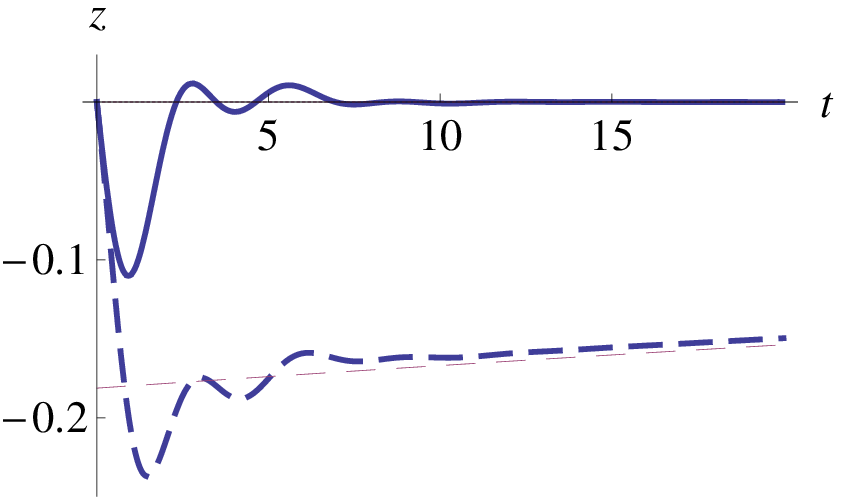}\\
\caption{The averages of $x,~y$ and $z$ [Eq. (\ref{xyz1})]for
$\phi=-.3\pi,~J^{}_d=.5,~\zeta^{}_d=.3,~q^{}_J=q^{}_\zeta=.5,~\zeta^{}_\Delta=0,~p^{}_+=p^{}_-=1/2$.
The initial qubit state [Eq. (\ref{psi})] is given by
$\alpha=\gamma=.25\pi$. All energies and inverse times are in
units of $\lambda$. The full (dashed) thick lines correspond to
$\Delta=0$ ($0.2$). The full (dashed) thin lines represent the
exact (approximate) asymptotic behavior. The derivation of the
dashed lines, for $\Delta\ne 0$, is described in Sec. V.}
\label{sym}
\end{figure}

\end{widetext}

To present our results graphically, we follow conventional
notations\cite{Allen} and write the reduced density matrix in the
form
\begin{align}
\rho\equiv ({\bf I}+{\bf r}\cdot\sig)/2\ ,
\end{align}
where the (real) Bloch vector ${\bf r}\equiv (x,~y,~z)$ is defined
by
\begin{align}\label{xyz1}
&\rho^{}_{11}\equiv (1+z)/2\ ,\ \ \rho^{}_{22}\equiv (1-z)/2\ ,\ \
\rho^{}_{12}\equiv (x-iy)/2\ .
\end{align}
The full thick lines in Fig. \ref{sym} show the time evolution of
the average components of the Bloch vector in the symmetric case,
for one example of the parameters. Indeed, both the real and the
imaginary parts of $\rho^{}_{12}$ approach finite limits, while
$z\rightarrow 0$. These limits, given by Eq. (\ref{r11}), are
shown by thin lines.

\begin{figure}[ht]
\includegraphics[width=7cm]{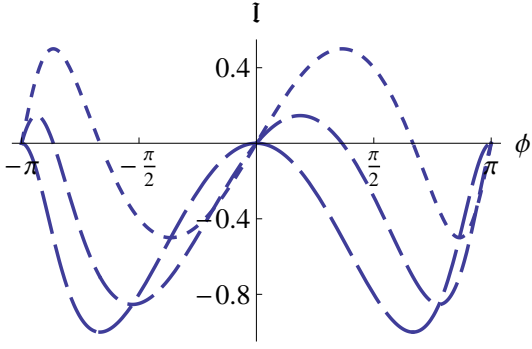}\
\caption{The flux dependence of the  asymptotic average current
(in units of $J^{}_0$) in the symmetric case, for
$q^{}_J=q^{}_\zeta=1/2$ and $\alpha=.25\pi$. Increasing dashes
correspond to $\gamma=0,~\pi/4$ and $\pi/2$. } \label{figcur}
\end{figure}

We now present more details on the derivation of Eq.
(\ref{curcur}). The general expression for the circulating current
is given in Eq. (\ref{netcur}). With the noise, this equation
gives the conditional averages, so that we need to average over
${\bf J}^{}_u\rh^{}_{21}$.   In the stationary state one has
$\rh^{}_{12}\rightarrow \rho^{}_{12}(\infty){\bf
p}^{}_0=e^{i\theta}{\rm Re}[e^{-i\theta}\rho^{}_{12}(0)]{\bf
p}^{}_0$ [see Eq. (\ref{r11})]. Substituting also ${\bf
J}^{}_u=J^{}_u{\bf I}+\zeta^{}_u\sig^{}_z$, and noting that the
average of $\sig^{}_z{\bf p}^{}_0$ is equal to $\Delta p$ and that
$J^{}_u$ and $\zeta^{}_u$ are real, one obtains Eq.
(\ref{curcur}). Figure \ref{figcur} shows the flux-dependence of
this asymptotic current, for $q^{}_J=q^{}_\zeta=1/2$ and several
values of the initial qubit relative phase $\gamma$ [Eq.
(\ref{psi})]. Interestingly, the current is odd (even) in $\phi$
for $\gamma=0$ ($\gamma=\pi/2$), but is neither odd nor even for
intermediate values of $\gamma$. Equation (\ref{curcur}) gives the
current in terms of the initial qubit parameters $\alpha$ and
$\gamma$. This current can in principle be measured by measuring
the orbital magnetic moment of the electron on the ring. To
retrieve $\alpha$ and $\gamma$, we need to perform three
preliminary measurements. Fixing the flux at a non-trivial value
$\phi^{}_1$ (not an integer multiple of $\pi$), one should measure
the asymptotic current for two known initial states. These
measurements determine the device parameters $\theta^{}_1$ and
$J^{}_0$. Repeating the same procedure for another flux
$\phi^{}_2$ and one known initial state, one would find
$\theta^{}_2$. The information on an unknown initial state can
then be extracted by  measuring the asymptotic current for the
same two fluxes.

\section{General case}

Since the calculations for the general case are somewhat
technical, we start with a brief summary of the results. Below we
present the full solution for the time evolution of the reduced
density matrix, which we have used to plot the dashed thick lines
in Fig. \ref{sym}. In this figure, as well as in much of the
discussion below, we concentrate on small deviations from the
symmetric case, namely
\begin{align}\label{small}
|\Delta \zeta^{}_J-J^{}_{12}\zeta^{}_\Delta|\ll \lambda,~\Omega\
,\ \ \ |q^{}_J-q^{}_\zeta|\ll 1\ ,
\end{align}
see Eqs. (\ref{cond1})-(\ref{qq}). As can be seen from the figure,
all three components of the Bloch vector [Eq. (\ref{xyz1})]
exhibit transient oscillations and then decay with a simple
exponential,
\begin{align}
\label{lxyz} x\approx x^{}_0e^{-t/\tau^{}_0},\ \ y\approx
y^{}_0e^{-t/\tau^{}_0},\ \ z\approx z^{}_0e^{-t/\tau^{}_0}\ .
\end{align}
To leading order in the small parameters (\ref{small}), the slow
decay rate is found to be
\begin{align}\label{tau00}
\tau_0^{-1}\approx \frac{(\zeta^{}_\Omega\Delta-\zeta^{}_\Delta
\Omega)^2+4J^{2}_{\cal
I}(\zeta_\Delta^2+\zeta_\Omega^2)}{\lambda[\lambda^2(\Omega+\zeta^{}_\Omega
\Delta p)^2+(\Omega^2-\zeta_\Omega^2)^2]}\ ,
\end{align}
where \begin{align}\label{para}
 &\Omega=\Delta^{}_{+-}=2{\rm
Re}[J^{}_{12}|\zeta^{}_J|/\zeta^{}_J]\
,\nonumber\\
&\zeta^{}_\Omega\equiv 2|\zeta^{}_J|\ ,\ \ \ J^{}_{\cal I}\equiv
(q^{}_\zeta-q^{}_J)\sin\theta\ . \end{align} Expanding also in
$\zeta^{}_\Omega$, this rate becomes
\begin{align}\label{tau1}
\tau_0^{-1}\approx \frac{(\zeta^{}_\Omega\Delta-\zeta^{}_\Delta
\Omega)^2+4J^{2}_{\cal
I}(\zeta_\Delta^2+\zeta_\Omega^2)}{2\Omega^2}S(\Omega)
\end{align}
[see Eq. (\ref{noisesp})]. Again, the relaxation time is related
to the spectral function of the noise. The corresponding decay
time, $\tau_0$, indeed becomes infinite in the symmetric limit
$\zeta^{}_\Omega\Delta/\Omega-\zeta^{}_\Delta=J^{}_{\cal I}=0$,
and remains very long for small symmetry breaking. This explains
the behavior observed in Fig. \ref{sym}. Unlike Eq. (\ref{tau}),
the power spectrum function $S$ now picks the Rabi frequency
$\Omega$ of the system, representing what Abragam calls the
`adiabatic modulation'.\cite{abragam} This emphasis on the Rabi
frequency is sometimes also called the `rotating wave
approximation'. The long relaxation time $\tau^{}_0$ can be
identified with the relaxational time $T^{}_1$, responsible for
the asymptotic decay of the diagonal elements of the density
matrix towards equal occupations (associated with the decay of $Z$
[Eq. (\ref{map})], which did not decay in the symmetric case).

\begin{figure}[ht]
\includegraphics[width=7cm]{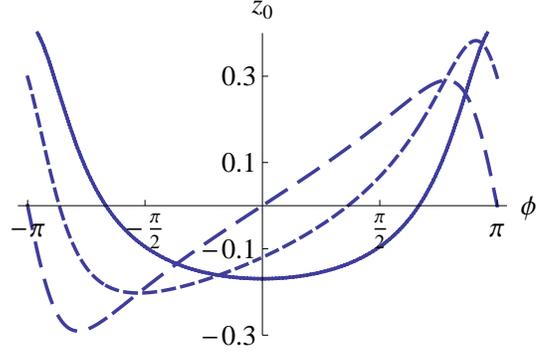}
\caption{The flux dependence of the amplitude $z^{}_0$ for the
same parameters as for the dashed lines in Fig. \ref{sym}, but
with $\gamma$ as in Fig. \ref{figcur}.} \label{zphi}
\end{figure}

Below we also evaluate the amplitudes $x^{}_0,~y^{}_0$ and
$z^{}_0$ [Eq. (\ref{lxyz})], and the approximate results are shown
by the thin dashed lines in Fig. \ref{sym}. The slow exponential
decay gives an opportunity to measure these amplitudes even after
a long time. Unlike the symmetric case, where asymptotically
$z\rightarrow 0$, one now finds non-zero values for the occupation
difference $z=\rho^{}_{11}-\rho^{}_{22}$. The flux dependence of
the amplitude $z^{}_0$ is shown in Fig. \ref{zphi}. As in Fig.
\ref{figcur}, note the even-odd dependence of $z(\phi)$. However,
all these coefficients can be tuned by a few preliminary
experiments done for a few initial qubit states and a few fluxes,
as discussed in connection with Fig. \ref{figcur}.

We now give more details. The unmotivated reader is welcome to
move to the next section.
 Since we mainly consider small deviations from the
symmetric case discussed in the previous section, we choose to
stay with the same basis used for that case, namely Eq.
(\ref{bab}). However, in order to decrease the number of
noise-related terms, we choose the phase $\theta$ as the phase of
$\zeta^{}_J$, so that $\zeta^{}_Je^{-i\theta}=|\zeta^{}_J|$
becomes real. This requirement is equivalent to Eq.
(\ref{tantheta}), upon replacing $q^{}_J$ with $q^{}_\zeta$ (in
the previous section we had $q^{}_J=q^{}_\zeta$). With the basis
(\ref{bab}), the two Hamiltonian operators become
\begin{align}\label{Hgen}
 &\mathcal{H}^{}_{\rm q}=(\Omega/2)(|+\rangle\langle +|-|-\rangle\langle
 -|)-(J^{}_{+-}|+\rangle\langle -|+{\rm h.c.})\ ,\nonumber\\
 &\mathcal{V}=(\zeta^{}_\Omega/2)(|+\rangle\langle +|-|-\rangle\langle -|)+(\zeta^{}_\Delta/2)(|+\rangle\langle -|+{\rm h.c.})\
 ,\nonumber\\
&J^{}_{+-}=-\Delta/2+iJ^{}_{\cal I}
\end{align}
 [see also Eq. (\ref{para})]. As a result, Eqs.
(\ref{eom2}) are now generalized to the form
 \begin{align}\label{neweom}
&(s{\bf I}-{\bf W})\widetilde{\bf
 Z}={\bf Z}(0)+2i({\bf J}^{}_{+-}\widetilde{\rh}^{}_{-+}-{\bf J}^{}_{-+}\widetilde{\rh}^{}_{+-})\ ,\nonumber\\
&(s{\bf I}-{\bf W}\pm
i\Omeg)\widetilde{\rh}^{}_{\pm\mp}={\rh}^{}_{\pm\mp}(0)\mp i{\bf
J}^{}_{\pm\mp} \widetilde{\bf Z}\ .
\end{align}
 Noting that \begin{align}
\widetilde{\bf P}{\bf p}^{}_0=\frac{1}{s+\lambda}\bigl
(1+\frac{\lambda}{s}{\bf T}){\bf p}^{}_0=\frac{1}{s}{\bf p}^{}_0\
,
\end{align}
 the first equation yields
\begin{align} \label{ZZs}
\widetilde{\bf Z}={\bf Z}(0)/s+2i\widetilde{\bf P}({\bf
J}^{}_{+-}\widetilde{\rh}^{}_{-+}-{\bf
J}^{}_{-+}\widetilde{\rh}^{}_{+-})\ .
\end{align}
Substituting this equation into the equations for
$\widetilde{\rh}^{}_{\pm\mp}$, yields two coupled equations for
the latter two 2-component vectors.

Solving these equations, and performing the inverse Laplace
transform, one finds that the time dependence of the off-diagonal
element of the density matrix $\rh^{}_{+-}(t)$ is a sum over
exponential terms, $e^{\alpha^{}_\ell t}$, where the
$\alpha^{}_\ell$'s are poles of $\widetilde{\rh}^{}_{+-}(s)$,
found as the roots of a sixth order real polynomial,
\begin{align}\label{pol}
d(s)=\sum_{\ell=0}^6~ d^{}_\ell s^\ell\ . \end{align} Substituting
the solutions for $\widetilde{\rh}^{}_{\pm\mp}$ into Eq.
(\ref{ZZs}) yields $Z(t)$. Equations (\ref{rho12}) are then used
to derive the time dependence of the average Bloch vector
$(x,~y,~z)$, as shown by the dashed lines in Fig. \ref{sym}.
Interestingly, there seem to be two main time scales. At the
beginning one observes a transient oscillatory behavior, up to
time scales of the order given by the symmetric case, Eq.
(\ref{alsym}). After that, all three variables exhibit a very slow
pure exponential decay, see Eq. (\ref{lxyz}).

 To explain this asymptotic behavior, we
return to the polynomial (\ref{pol}). The long-time limit of the
solutions is determined by the behavior of $d(s)$ at small Laplace
variable $s$, where we can use the approximation $d(s)\approx
d^{}_0+d^{}_1 s$. The Laplace transform then decays as
$e^{\alpha^{}_0 t}=e^{-t/\tau^{}_0}$, where
$\tau_0^{-1}=-\alpha^{}_0\approx d^{}_0/d^{}_1$. This
approximation is valid as long as $d^{}_0/d^{}_1$ is small. We
find
\begin{align} d^{}_0=
4\lambda^2 p^{}_+p^{}_-[(\zeta^{}_\Omega\Delta-\zeta^{}_\Delta
\Omega)^2+4J^{2}_{\cal I}(\zeta_\Delta^2+\zeta_\Omega^2)]\ .
\end{align}
Therefore, $\tau^{}_0$ becomes infinite, and one has a non-trivial
stationary solution, only when $d{}_0=0$, which happens only when
$J^{}_{\cal I}=0$ and either $\Delta=\zeta^{}_\Delta=0$ or
$\zeta^{}_\Delta/\Delta=\zeta^{}_\Omega/\Omega=|\zeta^{}_J|/{\rm
Re}[J^{}_{12}e^{-i\theta}]$, consistent with the results in Sec.
IV.  In all other cases, $\widetilde{\rh}^{}_{+-}$ and
$\widetilde{\bf Z}$ approach finite limits as $s\rightarrow 0$,
and therefore Eq. (\ref{lim}) implies that all  the components of
the Bloch vector approach zero as $t\rightarrow \infty$, leading
to the fully-mixed limit (\ref{mixed}).

 When the deviation from the symmetric
case is small, it is appropriate to expand the results in powers
of $(\zeta^{}_\Omega\Delta/\Omega-\zeta^{}_\Delta)$ and
$J^{}_{\cal I}/\Omega$. In our case, neglecting higher-order terms
in $\Delta,~\zeta^{}_\Delta $ and $J^{}_{\cal I}$, one has
$d^{}_1\approx \lambda[\lambda^2(\Omega+\zeta^{}_\Omega \Delta
p)^2+(\Omega^2-\zeta_\Omega^2)^2]$, yielding Eq. (\ref{tau00}).

 The coefficients $x^{}_0,~y^{}_0$ and $z^{}_0$ in Eq. (\ref{lxyz}) are the residues
 of the poles of the corresponding Laplace transforms at
 $s=\alpha^{}_0=-\tau^{-1}_0$. Therefore, to leading order in ${\bf J}^{}_{\pm\mp}$, these
 residues are the same as those for the pole at $s=0$ in
 the symmetric case. To this leading order,
 \begin{align}\label{x012}
 x^{}_0-i y^{}_0=e^{-i\theta}\sin(2\alpha)\cos(\theta+\gamma)
 \end{align}
and $z^{}_0=0$ [see Eq. (\ref{r11})]. Consequently, the amplitude
for the slow exponential decay of the circulating current is also
approximately given by Eq. (\ref{curcur}), as plotted in Fig.
\ref{figcur}. Measuring this amplitude therefore gives the same
information as discussed in connection with that equation (see the
end of Sec. IV.B).

Corrections to the next order in ${\bf J}^{}_{\pm\mp}$ just shift
the values of $x^{}_0$ and $y^{}_0$ slightly, and therefore we do
not discuss them here. In contrast, these corrections are crucial
for $z^{}_0=2{\rm Re}[\rho^{}_{+-}]$ [Eq. (\ref{rho12})], since
$z^{}_0=0$ at the zeroth order. To first order in ${\bf
J}^{}_{+-}=-(\Delta{\bf I}+\zeta^{}_\Delta\sig^{}_z)/2+i
J^{}_{\cal I}{\bf I}$, the second Eq. (\ref{neweom}) becomes
\begin{align}
\widetilde{\rh}^{}_{+-}=\widetilde{\bf
F}^{}_+[\rh^{}_{+-}(0)-i{\bf J}^{}_{+-}{\bf Z}(0)/s]\ .
\end{align}
The amplitude $z^{}_0$ is given by the average of $2{\rm
Re}[\lim^{}_{s\rightarrow 0}(s\widetilde{\rh}^{}_{+-})]$. Some
algebra then yields
\begin{align}\label{z0}
&z^{}_0\approx \lambda(c^{}_1\zeta^{}_\Delta+c^{}_2\Delta+c^{}_3J^{}_{\cal I})Z(0)/d^{}_1\ ,\nonumber\\
&c^{}_1=\lambda^2(\Omega+\zeta^{}_\Omega\Delta p)\Delta
p+(\Omega^2-\zeta^{2}_\Omega)(\Omega\Delta p-\zeta^{}_\Omega)\
,\nonumber\\
&c^{}_2=\lambda^2(\Omega+\zeta^{}_\Omega\Delta
p)+(\Omega^2-\zeta^{2}_\Omega)(\Omega-\zeta^{}_\Omega\Delta p)\ ,\nonumber\\
&c^{}_3=8 \lambda p^{}_+ p^{}_-\zeta^2_\Omega\ ,
\end{align}
where $Z(0)=\sin(2\alpha)\cos(\theta+\gamma)$. The thin dashed
lines in Fig. \ref{sym} were drawn using Eqs. (\ref{lxyz}),
(\ref{x012}) and (\ref{z0}). As can be seen, the approximation for
$z$ is excellent, while those for $x$ and $y$ are good apart from
a small shift which can be calculated from the next order. The
$\phi-$dependence of $z^{}_0$ is quite complicated, since the
coefficients $c^{}_\ell$ also depend on $\phi$, via $\Omega$ and
$\zeta^{}_\Omega$. When $q^{}_J\ne q^{}_\zeta$ then $J^{}_{\cal
I}$ is proportional to $\sin\theta$, introducing an additional
$\phi-$dependence.

\section{Discussion}

In this paper we discussed a qubit which is coupled to the
environment via a single telegraph noise variable $f(t)$. Apart
from the quantum information, which is stored in the qubit, the
system is characterized by the following parameters: the bare
energy gap $\Delta$, the bare hopping energies $J^{}_d$ and
$J^{}_u=J^{}_d q^{}_J$, and the amplitudes of the noise
$\zeta^{}_\Delta$, $\zeta^{}_d$  and $\zeta^{}_u=\zeta^{}_d
q^{}_\zeta$. To obtain partial decoherence we require
$\Delta=\zeta^{}_\Delta=0$ or
$\zeta^{}_\Delta/\Delta=\zeta/J^{}_0$ and $q^{}_J=q^{}_\zeta$. As
already mentioned, the energy gap $\Delta$ and the two hopping
energies $J^{}_u$ and $J^{}_d$ can all be tuned by gate voltages
on the two quantum dots and on the barriers along the hopping
paths. Therefore, one can in principle tune these parameters to
the partially decoherent limit.

Furthermore, $\zeta^{}_\Delta$ depends on the relative locations
of the noise source and the qubit. To reduce $\zeta^{}_\Delta$,
the noise source should be placed symmetrically relative to the
two dots. In that case, $\zeta^{}_1=\zeta^{}_2$, and therefore
$\zeta^{}_\Delta=0$. Also, if $\epsilon^{}_1$ and $\epsilon^{}_2$
represent two arbitrary levels of some large dot, they will not be
strongly affected if the volume of that dot is not sensitive to
the noise.
 For an arbitrary qubit state and an arbitrary flux,
we thus propose to tune $\Delta$ and $q^{}_J$ until one observes a
non-zero asymptotic circulating current, which also generates an
orbital magnetic moment. After such tuning one can use the same
system for retrieving the  quantum information, stored initially
on the qubit, from measuring the current and/or the magnetic
moment for any other flux and any other initial qubit state.

The measurement of equilibrium persistent currents is quite
difficult, and it is only recently that novel methods were
invented to measure them.\cite{persist} It remains to be seen if
such methods can also be applied to the circulating currents
discussed in the present paper. As mentioned, when the system's
parameters deviate from the special cases with partial decoherence
then one can also extract the initial qubit information from
measurements of the occupations of the states on the qubit's
quantum dots. Measuring the qubit dots occupations at real time is
difficult, since the measuring time should be much shorter than
any decoherence time. \cite{fuji} However, in the scheme presented
here one need not worry about the fast transient decay times, and
the long decay time $\tau^{}_0$ can in fact be tuned
experimentally. All one needs to do is tune the necessary gate
voltages and watch for a slow relaxation. Therefore, there is a
much better chance that existing methods for measuring dot
occupations will work here.

Equation (\ref{Vav}) suggests other options for `symmetric' cases.
For example, if $J^{}_{12}=\zeta^{}_J=0$ then
$z=\rho^{}_{11}-\rho^{}_{22}$ remains constant in time, while
$\rho^{}_{12}$ decays to zero. In this case, measuring the
time-independent dot occupations will yield $z(0)=\cos(2\alpha)$
[Eq. (\ref{psi})]. However, to gain the flexibility due to the
magnetic flux, and to extract information on $\gamma$, one would
still need to deviate slightly from this symmetric case. An
expansion in $J^{}_{12}$ and/or in $\zeta^{}_J$ would then yield
similar slow decays towards the fully-mixed state.

We expect similar qualitative results for more complex structures.
For example, one can replace each of the bonds $u$ and $d$ by a
path which goes via a linear chain of quantum dots, and one can
tune the energy level on one or more of these dots through a
resonance, thus changing the effective hopping energy
$J^{}_{u,d}$. As stated, we also expect similar results for other
sources of noise and for a system affected by more than one
fluctuator, provided one can tune the system to a `symmetric'
limit where the commutator $[\mathcal{H}^{}_{\rm q},\mathcal{V}]$
is small.

\vspace{.3cm}

\noindent {\bf Acknowledgements} \vspace{.2cm}

We acknowledge discussions with Y. Imry. SD is grateful to BGU and
to the Einstein center at the WIS for partially supporting his
visits to Israel.  AA and OEW acknowledge support from the DIP and
from the BSF, as well as discussions with P. Stamp and the
hospitality of the PITP at UBC.


\begin{thebibliography}{21}
\bibitem{sym}  An alternative way to solve the problem for a complex $J^{}_{12}$
 is to
  perform a gauge transformation from $|2\rangle$ to
  $|2'\rangle=e^{-i\theta}|2\rangle$. Then $J^{}_{12'}$ becomes
  real,
  the Hamiltonian becomes symmetric
 under $1 \leftrightarrow 2'$ and one can use the same solution as
 in Ref.$\ $\onlinecite{IT}. That is why we call this solution `symmetric'.
\end{thebibliography}
\end{document}